# Citizen Science:
# Contributions to Astronomy Research

CAROL CHRISTIAN
Space Telescope Science Institute
3700 San Martin Drive
Baltimore MD 21218, USA
carolc@stsci.edu

CHRIS LINTOTT
Oxford University
Department of Physics
Denys Wilkinson Building
Keble Road
Oxford OX1 3RH, UK
cjl@astro.ox.ac.uk

ARFON SMITH
Oxford University
Department of Physics
Denys Wilkinson Building
Keble Road
Oxford OX1 3RH, UK
arfon.smith@astro.ox.ac.uk

LUCY FORTSON
University of Minnesota
School of Physics and Astronomy
116 Church Street SE
Minneapolis MN 55455, USA
fortson@physics.umn.edu

AND
STEVEN BAMFORD
University of Nottingham
Centre for Astronomy and Particle Theory
School of Physics and Astronomy
University Park
Nottingham NG7 2RD, UK
steven.bamford@nottingham.ac.uk

**Abstract:** The contributions of everyday individuals to significant research
has grown dramatically beyond the early days of classical birdwatching and endeavors of
amateurs of the 19th century. Now people who are casually interested in science can
participate directly in research covering diverse scientific fields. Regarding astronomy,
volunteers, either as individuals or as networks of people, are involved in a variety of types
of studies. Citizen Science is intuitive, engaging, yet necessarily robust in its adoption of sci-
entific principles and methods. Herein, we discuss Citizen Science, focusing on fully
participatory projects such as Zooniverse (by several of the au-thors CL, AS, LF, SB), with mention of other programs. In particular, we make the case that citizen science (CS) can be an important aspect of the scientific data analysis pipelines provided to scientists by observatories.

**Introduction**

Over time, the term "citizen science" has changed. Previously, it was most often associated with the activities of individuals who contributed exten-sive observations, engaged in data collection related activities, and enjoyed camaraderie while producing both modest and significant contributions to the advancement of science. Currently, several permutations of the defi-nition of citizen science encompass not only a love of the natural world, but also other concepts such as the "democratization of science" including the formation of public policy by increasing public engagement in science. In this paper we refer to "citizen science" as the voluntary participation of individuals, many of whom may have no scientific training in scientific research. Their activities, such as visual observations, measurements, com-putations, or other investigations, contribute as an aggregate to original research goals. Specifically, the criteria used herein to define citizen science are:

- Non-expert participation: Scientific projects initiated by professionally trained researchers but participated in by volunteers without professional-level training in a relevant research area (and hence excluding self-funded researchers operating essentially as independent professionals).

- Many individuals: Involves networks of numerous individuals contributing to a specific set of goals, such that there are significantly more volunteers than professionals, usually by several orders of magnitude.

- Original Research: Research tasks that can significantly benefit from or are only possible through activities of a large number of humans. This last attribute is important when considering citizen science in the context of data processing and analysis.

For the purpose of this paper, we adopt a definition that emphasizes the active involvement of a large group of individuals to achieve real research objectives with measurable outcomes. Citizen scientists in this context are genuine collaborators in the resulting research, and in some cases, their activities are fundamental to studies of the physical universe as well as providing key components in improving automated data analysis and machine learning. For this paper, we highlight the relationship of citizen science to astronomical observatories and professional astronomy with the confidence that many of these principles broadly apply to other sciences.

We also caution that many activities and programs that are educational and may contribute to scientific research are not considered as citizen science in our context. For example, we are not discussing the "traditional" citizen activities carried out by amateur astronomers who discover supernovae, monitor variable stars and so on, because the extent and scope of these activities are beyond the purpose of this paper. Other examples that are not under our umbrella of citizen science include science projects specifically designed for students as educational exercises or activities that mimic or reproduce scientific results or methods, or individual student projects assisting professional researchers. We also will not address distributed computing activities that involve individuals, or more accurately, their computers along the lines of S*ETI@home*[1] in performing computational tasks for scientists.

---

[1] http://setiathome.berkeley.edu/sah about.php

This definition does, however, include a very wide range of projects. While many are simple classification tasks, complex scientific tasks also can be undertaken by citizen scientists. A well –developed, fairly involved example is Fold.it, a game-like experience in which participants develop new models of protein folding. The user can glean much scientific insight because opportunities for increasing understanding are cleverly embedded in the mechanics of gameplay, and a score that reflects the energy state of the protein is awarded[2].

**Zooniverse Citizen Science**

One of the largest collections of citizen science projects is the *Zooniverse*[3] that arose out of the *Galaxy Zoo* project, initiated in 2007 (Raddick et al. 2010). The idea behind *Galaxy Zoo* was to engage citizen scientists in examining almost a million galaxies contained within the Sloan Digital Sky Survey (SDSS) to create simple classifications for each object from their visual appearance as spiral or elliptical. These morphological classifications provide important statistical insights regarding the dynamical states, angular momenta, and star-formation properties of galaxies, which greatly aid our understanding of galaxy formation and evolution, the effects of environment, and fundamental cosmology. Automated classification algorithms can provide some information but, to date, the human eye and judgment remain superior analysis tools for morphology. Selecting galaxies by appearance rather than other characteristics such as color, brightness, etc. establishes a specific sample for different types of ensuing studies. Public response was extremely positive and ultimately *Galaxy Zoo* and *Galaxy Zoo 2* gathered over 60 million classifications. Several new research investigations have been spawned, and many papers published[4]

The *Zooniverse* has been expanded into other areas, so far mostly astronomy related but some beyond that discipline. The successor to *Galaxy Zoo 2* includes high resolution Hubble Space Telescope (HST) data from several programs including one of the observatory's *Treasury Programs*[5] to enable citizen scientist classifications of distant galaxies from a deep survey, and thereby contributing an additional dimension to galaxy morphology studies. *Galaxy Zoo: Hubble* is enabling direct examination of changes in the galaxy population from early in cosmic history until now.

Other astronomically related *Zooniverse* projects include the *Milky Way Project:* visual scrutinizing infrared images of the Milky Way galaxy to identify bubbles of material and contribute to the understanding of star formation environments and stellar collapse; *Solar Storm Watch:* spotting explosions on the Sun and tracking them across space to Earth, and *Galaxy Zoo Mergers:* selecting simulated models that best match images of real merging galaxies.

Thus, *Zooniverse* projects are based on human interpretation of complex phenomena through activities that cannot be thoroughly automated. This interaction with data is sometimes called *human computation*. Many *Zooniverse* projects provide tiers of training from simple to more complicated so citizen scientists can accomplish the tasks successfully and advance in expertise. *Zooniverse* projects, available through a web browser, are generally not dependent upon *when* or *where* the participant does the analysis. This contrasts with many other historical and current data collection activities that depend upon the location and timing of amateur contributions (cf. Budburst[6], Community Collaborative Rain, Snow and Hail[7] [CoCoRaHS], Coral Reef Alliance[8], eBird[9], Globe at Night Dark Skies Awareness[10], Christmas Bird[11]), and numerous others (including International Occultation Timing Association[12], Monarch Watch[13], SciSpy[14], and SKYWARN[15]).

---

[2]See Cooper et al. (2010), as well as Protein Folding at http://fold.it/portal/ and http://folding.stanford.edu/
[3]http://www.zooniverse.org
[4]See Lintott et al. (2009 & 2011), as well as a full set of references at http://www.galaxyzoo.org/published papers
[5]http://archive.stsci.edu/hst/tall.html

## Key Advantages – why citizen science is a significant component of large dataset analysis?

The kinds of citizen science activities employed by *Zooniverse* projects are primarily an attempt to deal with the flood of data being produced in astronomy, especially by increasingly automated large-scale surveys. Specifically, the *Zooniverse* allows a distributed community of volunteers to analyze professionally collected data.

**Scalability.** The first advantage of citizen science is its scalability. In a prescient paper, Lahav et al. (1995), discussing the then forthcoming Sloan Digital Sky Survey, noted that "Classifying very large data sets is obviously beyond the capability of a single person,". Distributing data to large numbers of individuals and collecting classification results for any given project could be a daunting management challenge, and so automated processes were seen as the most pragmatic solution.

Two circumstances have helped to greatly improve the practicality of citizen science. First, the adoption of public data sets and archives adhering to agreed-upon formats allows ready access to the data by a broad community. Second, the rapid development of communications technology in the last fifteen years, and resulting pervasion of the Internet in everyday life, enables very large numbers of volunteers to be engaged on a project. By dramatically scaling up the number of people available to work on a problem, citizen science eases the problem of the data flood by alleviating the reliance solely on automatic classifications, which often do not yet perform as well as the human techniques they were intended to replace.

Typically in the *Zooniverse,* such projects see a large spike in traffic just after activation, although sustained attention is achievable. For example, in the *Planet Hunters* project[16] originated in December 2010, participants look for transits in data from NASA's Kepler satellite. In May 2011, five months after launch, *Planet Hunters* was still attracting activity that amounts to the equivalent of 51 Full Time Equivalent staff[17]. *Galaxy Zoo* alone has recorded more than 150 million classifications over nearly four years.

---

[6] http://neoninc.org/budburst/

[7] http://www.cocorahs.org/

[8] http://coral.org/

[9] http://ebird.org/content/ebird/

[10] http://www.darkskiesawareness.org/gan.php

[11] )http://birds.audubon.org/christmas-bird-count

[12] http://www.occultations.org/

[13] http://www.monarchwatch.org/

[14] http://scispy.discovery.com/

[15] http://www.nws.noaa.gov/skywarn/

[16] http://www.planethunters.org

[17] 1 FTE = 2080 hours a year = 52 weeks @ 40 hours a week

This enormous effort would be largely wasted if it produced results that were not useful to scientists. Careful design and testing is required prior to project initiation in order to ensure that the time invested by citizen scientists is fruitful. To this end, many citizen science activities have standalone tutorials, and some (including the original *Galaxy Zoo* and NASA's *Stardust@home*, which asked participants to identify dust grains in videos of aerogel retrieved from the Stardust mission to Comet Wild-2) even insist that a test is completed before allowing participation. Instead, the *Zooniverse* adopts a particularly effective system by incorporating participant guidance into the classification process (see Fig. 1). Careful weighting of users (Lintott *et al.* 2008) can be used to identify volunteers who perform particularly well when compared to expert data. This has the critical advantage of allowing different weightings to be used on the same database of classifications, extracting different user behaviors *post hoc* as may be required. This is analogous to weighting various automated algorithms or tuning software based processes for the particular science being pursued.

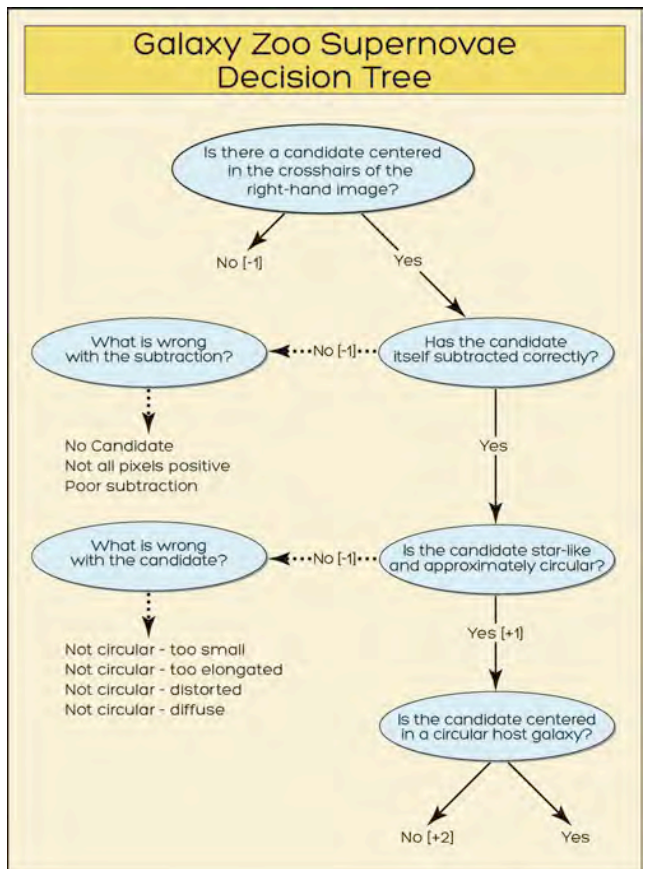

Figure 1. Decision Tree and scoring scheme for Galaxy Zoo Supernovae.

Further efficiencies can be achieved through a judicious choice of images to be considered. The *Galaxy Zoo Supernov*ae project[18] aims to rapidly assess candidate transients supplied by the Palomar Transient Factory (PTF)(Law *et al.* 2009). Once data is uploaded, subscribing volunteers receive an alert asking for their help in assigning a score to each candidate according to answers offered for simple questions, as illustrated in Figure 1 (adapted from Smith *et al.* 2011). This score not only calculates the final result, but also the priority with which candidates are presented for further inspection, ensuring that robust results can be obtained rapidly. This near real time intervention in data collection will be increasingly important as survey size continues to increase and rapid decisions need to be made as to what to follow up or even record, for example in transient object surveys.

---

[18]http://supernova.galaxyzoo.org

**Serendipity.** A second key advantage of human classification is that it preserves the opportunity for serendipitous discovery. Humans involved in a classification task will continue to look for unusual objects. Further, it has been seen that the interplay of the science team and citizen scientists can be multifaceted and can follow unexpected and quite fruitful trajectories that might be quite different than the original research goals. For example, *Galaxy Zoo* volunteer Hanny van Arkel spotted an unusual, and previously unknown, object, which became known as Hanny's Voorwerp[19]. This object was followed-up by studies using several ground-based observatories, the Swift satellite, and with HST by the Hubble Heritage Team[20]. These studies indicate that Hanny's Voorwerp is a unique example of a quasar light echo. A vast, twisted filament of gas is observed to be illuminated by the powerful beam of a quasar that existed in the center of IC 2497 until recently, but is no longer active. The HST campaign, in particular, suggested that gas outflow from the spiral galaxy IC 2497 is interacting with a nearby part of the filament, producing star formation.

Another remarkable unforeseen occurrence involved a self-organized group of Zooniverse volunteers who identified, and then systematically searched for members of a new class of compact galaxies, known as "Galaxy Zoo Peas" (Cardamone *et al.* 2009) for their characteristic strong green color in the SDSS images. These galaxies have star formation efficiencies more commonly associated with Lyman-break galaxies at redshifts of $z\sim2-3$. In this case, the citizen scientists were able to make significant progress in investigating the "Peas" using tools provided both by *Galaxy Zoo* and the SDSS itself. Building tools and systems that encourage this higher-level behavior is an important part of citizen science development as it offers both interest for advanced users and a valuable route to discovery for the citizen scientists as well as their professional collaborators.

**Relation to Machine Learning**. A third important advantage is the ability of citizen science projects to produce large training sets that can be used to improve machine learning approaches. As well as the benefits of size, there is some evidence (e.g. Banerji *et al.* 2010) that the richness of the training sets, which include not only definitive answers but also quantitative estimates of difficulty of classification, can lead to significant improvement in automated techniques.

Tasks can be divided between machine and citizen to take advantage of their different capabilities. The supernova project discussed above is a simple example, in which the PTF's pipeline is responsible for identifying potential supernovae that are then assessed by volunteers. In the future, these two modes will become more integrated in real time, with observatory pipelines deciding which data to pass to humans based on an assessment of the data being received. For example, training sets could be generated on the fly hour-by-hour as sky conditions and instrument performance changes or the proportion of candidates passed to humans could vary as the number of volunteers waxes and wanes over the course of a day. Alternatively unusual images could be flagged for human follow-up according to some preset criteria. Extensive development of methods to balance decisions made by several different machine learning techniques has already taken place, and should be easily adapted to include citizen scientists as a classification option.

---

[19] See Lintott et al. (2009), as well as The Galaxy Zoo and Hanny's Voorwerp (20at http://www.ing.iac.es/PR/SH/SH2008/zoo.html
[20] http://hubblesite.org/newscenter/archive/releases/2011/01/

This vision, in which appropriate citizen science projects could form a natural part of an observatory's data reduction pipeline is ambitious, and demands some agility and adaptability on the part of observatory facility scientists and engineers, but the key advantages of citizen science outlined above make it imperative if we are to continue to make the most of the data available.

One of many methods for measuring the success of citizen science in terms of research productivity is the suite of case studies of serendipitous discoveries that came about due to the activities of the volunteers, mentioned above. Other measures that are commonly used to access science productivity relate to publications. Consider that *Galaxy Zoo* has only been in operation for approximately 3 years; it is remarkable that 21 refereed papers and nearly 500 citations have resulted from the citizen science work. In addition, three of the top 100 SDSS papers in the last three years are from *Galaxy Zoo*. The citizen science research, while small, has contributed to about 2-3% of the SDSS publications to date (although SDSS publications have amassed over a considerably longer period of time). Citizen science activities have a great potential to grow and contribute more significant results to astronomical research.

## Techniques, Tools and Infrastructure – How to make CS happen?

Distributed data analysis of the kind discussed in this chapter is only possible thanks to the widespread availability of the World Wide Web. The fate of the original *Galaxy Zoo*, which was overwhelmed by demand for much of its first day, illustrates the need to choose tools appropriate to accommodate the spikes in traffic received by many citizen science projects.

Commercial cloud computing delivers a solution for web applications with unpredictable traffic. Load balancing and auto-scaling tools supplied by brokers such as Amazon Web Services ensure that new servers can rapidly and automatically be made available whenever necessary. A modular implementation of the application, in which a database and application layer communicates with a thin website layer *via* a RESTful Application Programming Interface (API) makes this scaling easier, and helps make code reuse between projects more viable. This structure can also support alternative presentations of an interface, with encouraging results. *Tinyplanets.com* (a commercially developed educational site for 5-7 year olds) hosts a version of *Zooniverse's Moon Zoo* interface; Warner Brothers have developed a version of the *Milky Way Project* to tie in with their 'Green Lantern' film; and the *Galaxy Zoo* iPhone app has seen an increase of two orders of magnitude compared to the web browser version in the number of classifications *per user*; impressive, even given the likely selection effects in participation that are operating.

In addition to the main site, tools which encourage communication between the volunteers and the project scientists and developers are of vital importance in attracting and sustaining a community. Much of the serendipitous science from *Galaxy Zoo* came from a basic forum, and a new 'Talk' tool that can be more closely integrated with the process of classification itself has been developed and released[21]. The ultimate goal of such tools should be to bring questions and interesting discoveries to the scientists' attention only when expert input is necessary, reducing the time needed for appropriate mentoring while still ensuring nothing gets lost. For more one-way communication, most citizen science projects use blogs to keep the community informed about the progress of research.

---

[21] https://github.com/zooniverse/Talk

**Citizen Science and Observatories**

Astronomical observatories, especially those with federal funding, provide access to facilities including observational equipment and data capture primarily for scientific research. In order to maximize the science return on the funding investment, open data policies are implemented, delivering access to archives for a wide community. In addition, it is the practice of current observatories to implement facility pipeline processing for calibration and instrument characteristic removal, populating the archives not only with raw observations but higher-level science data. Experience has shown that observatories that do accommodate robust archives and data analysis pipelines and tools enjoy remarkable productivity (Apai *et al.* 2010).

Traditionally, there has not been any routine relationship between professional observatories and citizen science. There have been isolated examples of public participation and the integration of amateur activities into science research, but observatories have not adopted citizen science contributions as routine business. With the growing emphasis on large, complicated datasets, as well as the acquisition of survey data that cover a significant portion of the sky, additional analysis resources may be required. Experience with the *Zooniverse* has demonstrated that citizen science can make a contribution. Thus, the addition of a citizen science infrastructure to the data analysis pipeline may be valuable and, in fact, advisable, in order to improve science productivity for specific types of data and research problems. For example, as shown in Fig. 2, data flows from the observatory to the systems that provide data quality checking, calibration, data analysis pipeline processing and the production of higher-level science data products. Output data and products are deposited in the observatory archive and can be delivered to a citizen science project. The citizen science "pipeline" produces additional information that can be merged with the other data products, and stored in the archive.

Of course it is possible for curators of large private datasets also to make use of citizen science as an additional analysis tool, but one should note that a by-product of restricted data access policies is that the public user (as well as members of the science community) become disinterested in the data. This creates a situation where many talented analyzers/scientists (unaffiliated with the data) do not spend time contributing to challenging research studies related to the proprietary data. This sort of restriction also has a dampening effect on the potential for serendipitous and spontaneous citizen-led discovery and research.

In order for a research topic to benefit from unique human cognitive abilities exercised by a collective of many individuals it must engage and challenge the participants by ensuring that they work with real scientific data, information, and resources. Citizen science pipelines must be designed with care to be the most beneficial for both the science teams and the citizen scientist. The implementation of citizen science contributions necessarily involve the scientists with a vested interest in the research goal. *Zooniverse* experience demonstrates that citizen participants are highly motivated not only by contributing to research, but also by having a connection to the research team.

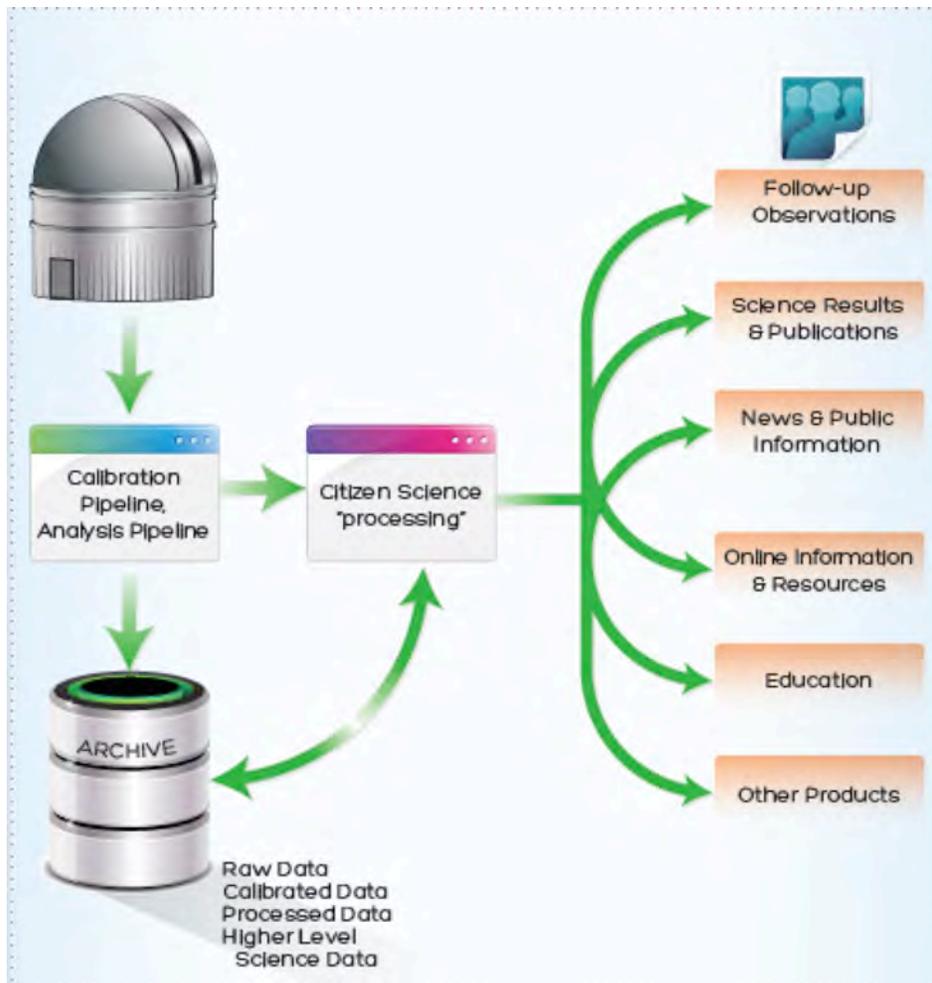

Figure 2: Notional flow of data processing for astronomical observations, including standard pipelines, archive, and Citizen Science analysis pipeline. (@STScI)

It is clear that citizen science volunteers need appropriate infrastructure and tools. Access to imagery and visualization of sky locations can also be an engagement tool, promoting a sense of ownership for the individual, even if the analysis is of some other sort of data product (for example, a light curve). Some individuals will accomplish further investigations, confer with other citizen scientists, and will report back on their individual work, if the appropriate forums are available. This does imply that the research team or the observatory must make a modest investment in citizen science infrastructure, services, and tools to achieve success.

To maximize the contribution of the "citizen science module" (Fig. 2) in the data analysis pipeline, individuals must find the activities in several ways. For example, the *Galaxy Zoo Hubble* project created as a follow-on to the SDSS-based *Galaxy Zoo* was advertised through both the *Zooniverse* and HST's HubbleSite.org. This latter site is the main public interface to HST and has myriad resources including activities, information, and interactive modules.

The majority of citizen scientists participated through the *Zooniverse,* but the clientele of HubbleSite also participated. Thus coordination between the main citizen science portal and the observatory portal was fruitful and cast a wider net for participation.

## Summary

Citizen Science, as we use the term in this chapter, provides a professionalized manner in which any individual can contribute to substantive, authentic scientific research. In particular, the *Zooniverse* projects have demonstrated that research projects can significantly benefit from large numbers of participants in cases especially where human cognitive abilities can supplement automated data analysis.

Initial results have shown that for observatories collecting large, sometimes complicated and also survey type datasets, *Zooniverse* methodology produces robust results as well as serendipitous discoveries. Specifically, citizen scientists have contributed to the results from the large SDSS sky survey, the concentrated transient/planet finding studies from the NASA Kepler mission, characterization of lunar craters and features from the Lunar Reconnaissance Observatory, and the galaxy morphology studies from HST Treasury programs, to name a few.

Selection of projects is critical if we are not to waste the time of volunteers or to fail to meet the goal of providing authentic engagement with research. Basic data analysis task should, where possible, be automated rather than thoughtlessly passed to citizen scientists. Instead, by incorporating the potential to support citizen science into a standard pipeline as illustrated in Fig. 2, it should be possible to allow scientists with a need to use the resources of volunteers to rapidly prototype, test and launch a project at whatever level is necessary.

The presence of an existing community of engaged citizen scientists also allows for the possibility of 'career' development for the citizen scientists themselves. As discussed above, engagement is possible at many levels of task from simple classification, to investigation of interesting objects and eventually to follow-up or detailed work. This progression to higher level tasks will be a necessary part of adapting citizen science for the long term as automation of basic data analysis continues, driven in part by the results of the citizen science projects themselves.

Reflection upon the growth and success of this type of citizen science has led us to believe that it could be quite beneficial to integrate citizen science analysis "modules" (projects) into an observatory pipeline in some form. Citizen science will continue to grow and evolve along its own trajectory. Observatory personnel would be wise to monitor citizen science techniques for possible adaptation to standard high level data processing. Certainly when new sky survey observatories are planned, citizen science contributions should be considered from the outset. One instance is the planning for the Large Synoptic Survey Telescope (LSST[22]). While citizen science is not yet fully integrated into the data analysis pipeline, it is planned that data from the LSST pipeline will be available for citizen science projects.

It also has not escaped our notice that astronomical citizen science makes a significant contribution to public understanding of science, with a growing potential for formal and informal science education. Our focus here, however, is the significant impact that citizen science is having on scientific research.

---

[22]http://www.lsst.org/lsst

**Acknowledgements**

This work was supported by the NASA contract, NAS5-26555, to the Association of Universities for Research in Astronomy, Inc. for the operation of the Hubble Space Telescope at Space Telescope Science Institute. Additional support from NSF CDI grant DRL0941610 and ISE grant DRL0917608 and a Science and a Technology Facilities Council Advanced Fellowship. Gratitude is expressed to all the Zooniverse participants for their work and to the support from the STScI Office of Public Outreach staff.